\documentclass[twocolumn,aps,prc,superscriptaddress,showpacs]{revtex4}
\usepackage{graphicx}  
\usepackage{dcolumn}   
\usepackage{bm}        
\usepackage{verbatim}   
\usepackage{CJK}
\usepackage{amsmath,amssymb,amsfonts,amsthm}

\newcommand{\degree}{^\circ}

\begin{document}

\title{Yield ratio of neutrons to protons  in $^{12}C(d,n)^{13}$N  and $^{12}C(d,p)^{13}$C from 0.6 MeV to 3 MeV}

\author{W. J. Li}
\affiliation{Shanghai Institute of Applied Physics, Chinese Academy of Sciences, Shanghai 201800, China}
\affiliation{University of the Chinese Academy of Sciences, Beijing 100080, China}

\author{Y. G. Ma \footnote{Corresponding author: ygma@sinap.ac.cn}} 
\affiliation{Key Laboratory of Nuclear Physics and Ion-Beam Application (MOE), Institute of Modern Physics, Fudan University, Shanghai 200433, China}
\affiliation{Shanghai Institute of Applied Physics, Chinese Academy of Sciences, Shanghai 201800, China}

\author{G. Q. Zhang \footnote{Corresponding author: zhangguoqiang@zjlab.org.cn}}
\affiliation{Shanghai Advanced Research Institute, Chinese Academy of Sciences, Shanghai 201210, China}
\affiliation{Shanghai Institute of Applied Physics, Chinese Academy of Sciences, Shanghai 201800, China}

\author{X. G. Deng}
\affiliation{Key Laboratory of Nuclear Physics and Ion-Beam Application (MOE), Institute of Modern Physics, Fudan University, Shanghai 200433, China}
\affiliation{Shanghai Institute of Applied Physics, Chinese Academy of Sciences, Shanghai 201800, China}

\author{M. R. Huang}
\affiliation{
College of Physics and Electronics information, Inner Mongolia University for Nationalities, Tongliao, 028000, China}
\author{ A. Bonasera}
\affiliation{
Cyclotron Institute, Texas A\&M University, College Station, Texas 77843, USA.}
\affiliation{
Laboratori Nazionali del Sud, INFN, via Santa Sofia, 62, 95123 Catania, Italy}
\author{D. Q. Fang}
\affiliation{Key Laboratory of Nuclear Physics and Ion-Beam Application (MOE), Institute of Modern Physics, Fudan University, Shanghai 200433, China}

\author{J. Q. Cao}
\affiliation{Shanghai Institute of Applied Physics, Chinese Academy of Sciences, Shanghai 201800, China}
\author{Q. Deng}
\affiliation{Shanghai Institute of Applied Physics, Chinese Academy of Sciences, Shanghai 201800, China}
\author{Y. Q. Wang}
\affiliation{Shanghai Institute of Applied Physics, Chinese Academy of Sciences, Shanghai 201800, China}
\author{Q. T. Lei}
\affiliation{Shanghai Institute of Applied Physics, Chinese Academy of Sciences, Shanghai 201800, China}

\begin{abstract}
The neutron yield  in $^{12}$C(d,n)$^{13}$N and the proton yield in $^{12}C(d,p)^{13}$C have been measured by deuteron beam from  0.6 MeV to 3 MeV which is delivered  from a 4-MeV electro static accelerator  bombarding on the thick carbon target. The neutrons are detected at $0\degree$, $24\degree$, $48\degree$ and the protons at $135\degree$ in the lab frame. The ratios of the neutron yield to the proton one have been calculated and can be used as an effective probe to pin down the resonances. The resonances are found at 1.4 MeV, 1.7 MeV, 2.5 MeV in $^{12}C(d,p)^{13}$C and at 1.6 MeV, 2.7 MeV in $^{12}$C(d,n)$^{13}$N.  This method provides a way to reduce the systematic uncertainty and helps to confirm more resonances in compound nuclei.

\end{abstract}
\maketitle

\section{Introduction}

The reaction $^{12}$C(p,$\gamma$)$^{13}$N is drawing much attention both for its usefulness as an important reaction for studying the properties of light nuclei and for the significant role which is expected to play in the cores of main sequence stars and in the shells of giant stars \cite{Joseph}. The research of the $^{12}$C(p,$\gamma$)$^{13}$N reaction is a perspective to understand the well-known carbon-nitrogen-oxygen cycle (CNO cycle). The reactions occurring in the CNO cycle  \cite{brown_experimental_1962} are in the chain, 
$
    \notag^{12}C(p,\gamma)^{13}N(\beta^{+})^{13}C(p,\gamma)^{14}N (p,\gamma)^{15}O 
    \notag(\beta^{+})^{15}N(p,\alpha)^{12}C.
$
The result of the $^{12}$C(d,n)$^{13}$N reaction can be used in the research of the $^{12}$C(p,$\gamma$)$^{13}$N reaction with the Trojan Horse Method (THM)  \cite{typel_theory_2003}.
Compared with $^{12}$C(p,$\gamma$)$^{13}$N, $^{12}$C(d,n)$^{13}$N has 3 order higher cross section with incident energy around MeV, which makes it a better alternative method to determine the cross section for $^{12}$C(p,$\gamma$)$^{13}$N. In addition, the $^{12}$C(d,n)$^{13}$N reaction can be applied for the analysis of various of materials. It can also be used as a source of polarized neutrons \cite{jaszczak_c_1969}. The $^{12}C(d,p)^{13}$C reaction can be employed in metallurgy and Ion Beam Analysis (IBA) \cite{papillon_analytical_1997}. For example, the properties of steel are dependent on their carbon content and its spatial distribution. The cross section and the resonances of the nuclear reaction help greatly to understand the structure of the materials at a micro level  \cite{Tang,Peng}. The last but not the least, the deuterated polyethylene (CD$_{2}$)$_{n}$, is usually taken as the typical target for nuclear reactions induced by intensive lasers \cite{curtis_micro-scale_2018,Zhang2019}.  At very high plasma temperature, besides the  d(d,n)$^{3}$He reaction, the reaction $^{12}C(d,n)^{13}$N may happen. Since both reactions emit neutrons, which may result in a fake enhancement for d(d,n)$^{3}$He and make the measurement complicated. The cross section measurement for $^{12}C(d,n)^{13}$N will help to clarify the real cross section of d(d,n)$^{3}$He \cite{NuclTech1,NuclTech2,NuclTech3} in the laser induced plasma.

The cross sections of the $^{12}$C(d,n)$^{13}$N reaction and the $^{12}C(d,p)^{13}$C reaction have been measured in various of aspects. The cross section of $^{12}$C(d,n)$^{13}$N reaction is measured  by neutrons detection with a 4$\pi$ detector \cite{jaszczak_c_1969}. The cross section of $^{12}$C(d,n)$^{13}$N reaction is measured by counting the $\gamma$-ray from $^{13}N$ produced in the  reaction   \cite{michelnn_excitation_nodate}. The cross section of $^{12}C(d,p)^{13}$C reaction is measured by counting the $\gamma$-ray \cite{papillon_analytical_1997} . Though much effort has been paid, there exists a great diversity for the resonances and the cross sections at around 1.4 MeV among the data.
In this work, we  have conducted an experiment to study the ratio between the neutron yield in $^{12}$C(d,n)$^{13}$N  and the proton yield in $^{12}C(d,p)^{13}$C. This ratio can not only give the information of the resonances in the two reactions, but also eliminate the systematic error caused by the uncertain incident beam's intensity and can be applied in thick target reaction systems. 

\section{Experiment}
\label{sec:Experiment}
The experiment was performed in the Shanghai Institute of Applied Physics, Chinese Academy of Sciences. In this experiment, the deuteron beam with energies from 0.6 MeV to 3 MeV was used to bombard thick graphite target. Two reactions were supposed to be observed in the experiment, $^{12}$C(d,n)$^{13}$N and $^{12}C(d,p)^{13}$C. The ratio of the yields of neutrons to protons is extracted.

\subsection{The layout of the experiment}
\label{sec:The layout of the experiment}

\begin{figure}
  \centering
  \includegraphics[scale=0.6]{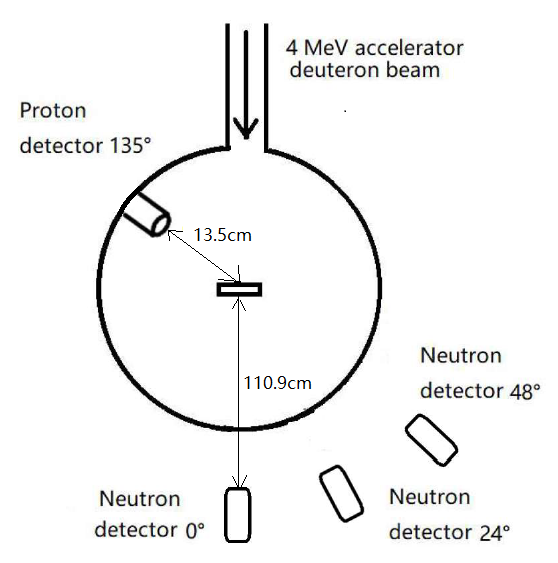}\\
  \caption{The layout of the accelerator, the detectors and the target.}
  \label{geometical arrangement}
\end{figure}

Fig.~\ref{geometical arrangement} shows the layout of the devices used in the experiment. The target chamber was a cylinder with diameter 60.0 cm. The beam tube leds the deuterium beam from the accelerator to the chamber. Three neutron detectors were put outside the chamber. The carbon target, the Faraday's cup and a proton detector were put inside the chamber.

The deuteron beam in the experiment was produced from the 4-MeV electrostatic accelerator. The voltage of accelerator reached 3 MeV in this experiment. The fluctuation of the voltage was lower than 2 keV, which guarantees a good energy resolution for the beam. The beam intensity was about several nA. There were three ion beam analysis terminals. The experiment was performed at the $15\degree$ terminal. This terminal provided conventional ion beam analysis. The vacuum reached $6.2\times 10^{-5} Pa$.

The graphite target was put in the middle of the chamber, facing directly to the deuteron beam. The thickness of our target was 1 mm which was much greater than the deuteron range, about several $\mu m$, in carbon. The abundance of the $^{12}C$ in the nature graphite in the experiment was 98.93\%.

A Faraday-cup, installed around the target in the chamber, was used to count the incident charged particles. However, the conductivity of thick graphite target is not so good that sparking during the experiment was unavoidable, which made the incident beam's intensity became hard to estimate. Our method, calculating and analyzing the ratio of the neutron yield to the proton yield, gave us a solution to research the resonances in the two reactions even without precise information of the beam intense.

\subsection{Detectors}
\label{sec:Detectors}

\begin{figure}
  \centering

  \includegraphics[scale=0.45]{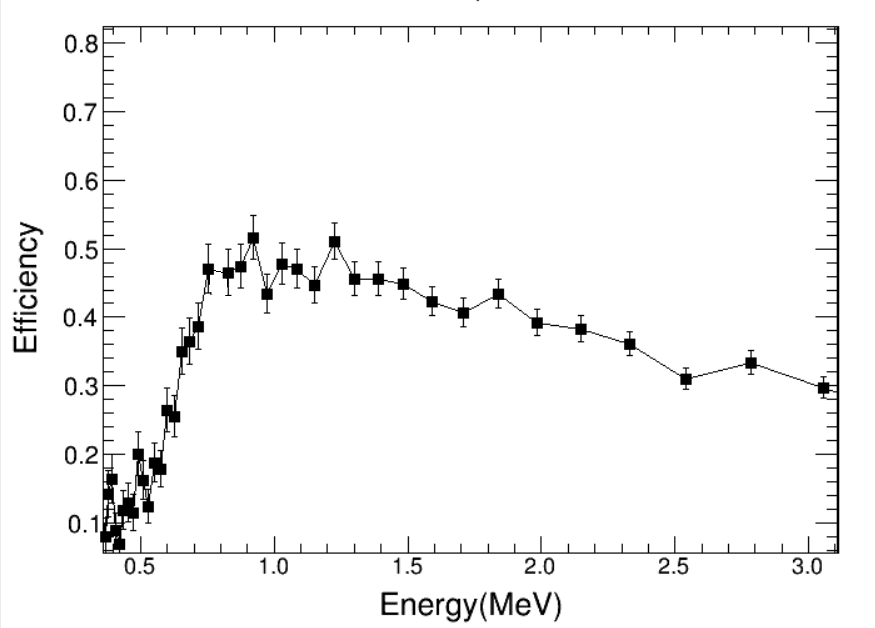}\\
  \caption{The efficiency of the EJ-301 neutron detector at $24\degree$.}
  \label{neutron efficiency}
\end{figure}

In the experiment, four detectors were used to count the yield, including three EJ-301 detectors for neutrons and an Au-Si surface barrier detector for protons.
The EJ-301 detector was a liquid detector which was suitable for the MeV neutron detection with a good neutron-gamma discrimination. Three EJ-301 detectors were put 112.9 cm away from the target at 0$\degree$, 24$\degree$ and 48$\degree$ from the beam axis in the experiment. The diameter of the EJ-301 detector was 12.7 cm. The solid angles of the three detectors were $1.2\times10^{-2}sr$.
A time of fly (TOF) system based on a Cf-252 neutron source was set up to calibrate the EJ-301 detectors.  The efficiency of the EJ-301 detector at 24$\degree$ varied from different incident energies was showed in Fig.~\ref{neutron efficiency}. The efficiency of the EJ-301 detectors was about 0.4 to 0.5 from 0.6 to 3 MeV. The final neutron yields were then corrected by this efficiency.
	
The Au-Si surface barrier detector was put 135 mm away from the target at 135$\degree$ from the beam axis. The diameter of the Au-Si surface barrier detector was 7.5 mm. The solid angle of the Au-Si surface barrier detector was $2.42\times 10^{-3} sr$. The efficiency of the Au-Si surface barrier detector was 100\% for the proton detection.

\subsection{Methods}
\label{sec:Methods}

A thick target with 1 mm thickness which is much larger than the range of deuteron in carbon is used in this work. The formula for yield of the nuclear reaction is as followed:
\begin{equation}
 Y=\int_{0}^{T}\int_{0}^{D}{I(E)N_{v}\sigma(E)dxdt},
\end{equation}
where D is the thickness, I(E) is the incident current and T is time duration of the projectile in the target. $N_{v}$ is the atom number per unit volume of the carbon target. $\sigma$(E) is the cross section of the reaction at energy E. The ratio of the neutron yield to the proton one in the two reactions is expressed as followed:
\begin{equation}
\frac{Y_n}{Y_p}=\frac{t\int_{0}^{D}{I(E)N_{v}\sigma_n(E)dx}}{t\int_{0}^{D}{I(E)N_{v}\sigma_p(E)dx}}
\end{equation}

Theoretically, a thick target is equivalent to the sum of several continuous thin targets. For instant, the yield with incident energy 3 MeV could be described by the formula as followed.
 \begin{equation}
Y_{total}=\bigtriangleup{Y_{0.6}}+\bigtriangleup{Y_{0.7}}+\cdots+\bigtriangleup{Y_{3.0}}
\end{equation}
The thickness of each thin target is the range that the deuteron cross through in the target from energy E to energy E - 0.1 MeV. $\triangle{Y_{E}}$ is the yield of the thin target. The total yield is the sum of the yields of all the thin target slices. For the two reactions in our work, when the incident energy is below 0.6 MeV, the cross section is less than 10mb. Therefore the slices in which the deuteron energy below 0.6 MeV will not be considered. For a very thin target, the formula for yield and cross section of the nuclear reaction is as followed:
 \begin{equation}
\triangle{Y}=I\triangle{t}N_v\sigma(E)\triangle{x}
\end{equation}
The yield in the thin target is proportion to the cross section.
The difference between the ratio of the neutron yield to the proton one in the two reactions is:
\begin{equation}
\left\{
             \begin{array}{lr}
            Y_{n_{E+0.1}}=Y_{n_{E}}+\triangle{Y_{n_{E+0.1}}}\\
            \\
             \frac{Y_{n_{E+0.1}}}{Y_{p_{E+0.1}}}-\frac{Y_{n_{E}}}{Y_{p_{E}}}=
             \frac{\frac{\triangle{Y_{n_{E+0.1}}}}{Y_{n_{E}}}-\frac{\triangle{Y_{p_{E+0.1}}}}{Y_{p_{E}}}}  {\frac{Y_{p_E}}{Y_{n_E}}    }\\

             \end{array}
\right.
\end{equation}

The resonances in the reactions can be identified by an extremely high cross section at a certain energy.
If a resonance occurs in the reaction at incident energy E, the yield of thin target $\triangle{Y_{N_E}}$ or $\triangle{Y_{P_E}}$ will have a rapid change, which may cause a maximum or minimum value in the ratio. The $\triangle{Y}$ is proportion to the cross section. Therefore, the resonances of the two reactions can be identified by the ratios.

\section{Results}
\label{sec:res}

\begin{figure}

  \includegraphics[scale=0.3]{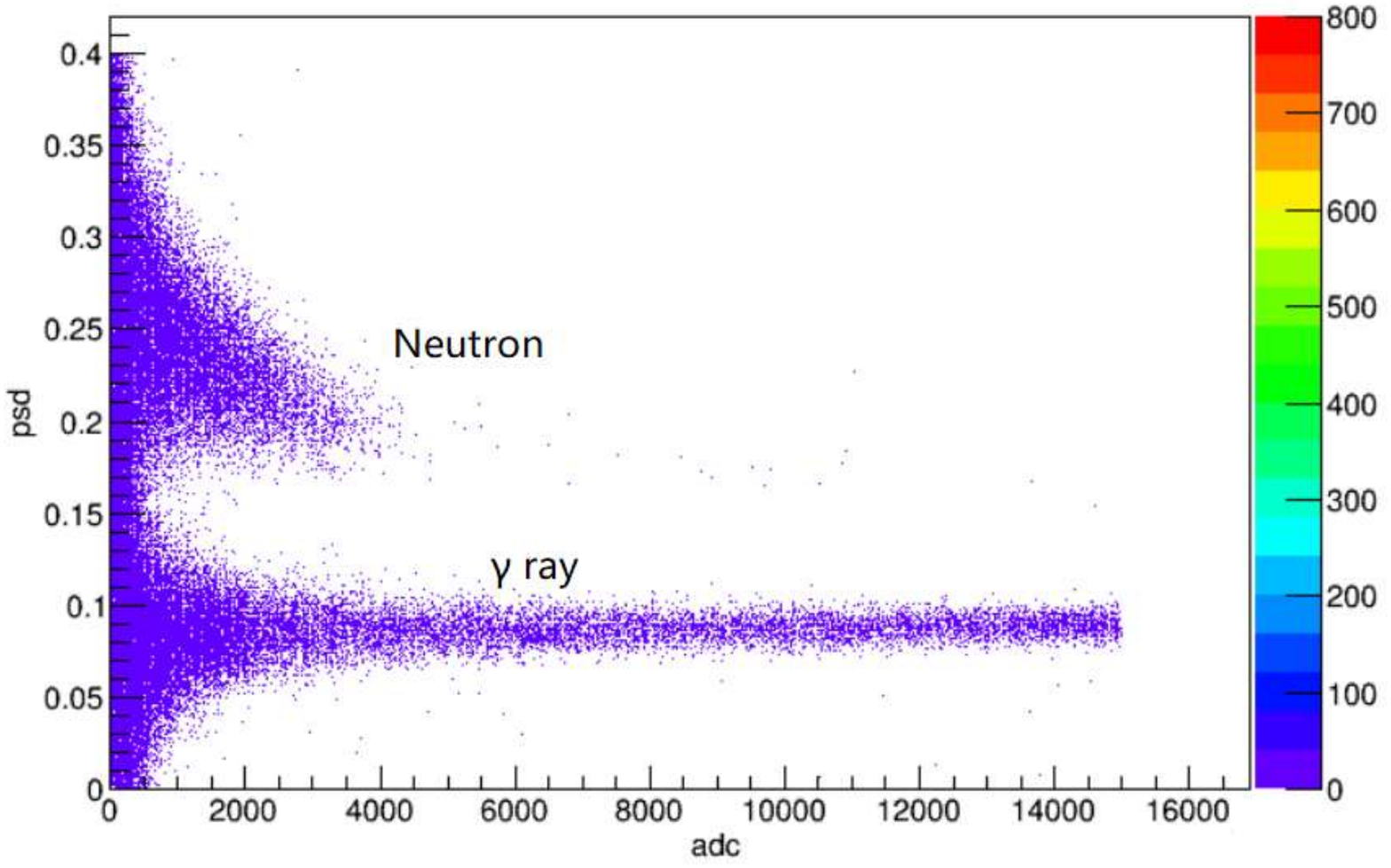}\\
  \caption{(Color Online) The signals of neutron detector at $48\degree$, incident energy 2 MeV.}
  \label{neutron detector}
\end{figure}

\begin{table*}[t]
$$
\begin{array}{|c|c|c|c||c|c|c|c|}
\hline
Incident Energy   &  Time   & Current  & Charge &Incident Energy   &  Time   & Current  & Charge\\
(MeV)&(s)&(nA)&(nC)&(MeV)&(s)&(nA)&(nC)\\
\hline
  0.600& 430 & 1.50  & 676 &1.900  & 300  & 0.12  &  36 \\
0.700  & 312 & 1.30  & 387 &2.000  & 300  & 0.15  & 77  \\
 0.800 & 296  & 1.00  & 300  &2.100  & 300 & 0.10 & 59 \\
0.900 & 300  & 1.00  & 295  &2.200  & 300 & 0.10  &  59 \\
 1.000 & 300  &  1.00 & 297 &2.300  & 300  & 0.10  &  64\\
1.100  & 300  & 0.70  &  198 &2.400  & 300  & 0.05  &  43\\
1.200  & 300  & 0.50  & 175 & 2.500  & 300  & 0.07  &  51 \\
1.300  & 300  & 0.40  & 120&2.600  & 300  & 0.10  &  92 \\
1.400  & 300 & 0.40  &  120 &2.700  & 300  & 0.08  &  54\\
1.500  & 300  & 0.20  & 125 &2.800  & 300  & 0.07  &  51  \\
1.600  & 300  & 0.20  &  70 &2.900  & 300  & 0.07  &  51 \\
1.700  & 300  & 0.20  &  70 &3.000  & 300  & 0.07  &  51 \\
1.800  & 300  & 0.20  &   48&&&&\\

\hline
\end{array}					
$$

\caption{The incident energy, the measuring time, the current intensity and the  aggregated charge in the experiment}
\label{layout}
\end{table*}
Table.\ref{layout} shows the deuteron beam current intensity, the total charge, the measuring time and the incident energy in this work. The measure time of each run is 5 minutes expect for the first three runs. A two-hour background measurement was also performed on the EJ-301 detectors.
Fig.\ref{neutron detector} shows an example of the signals of neutron detector at $48\degree$ when the incident energy is 2 MeV. The pulse shape discrimination (PSD) is calculated to distinguish neutrons and $\gamma$-ray. PSD is defined by:
\begin{equation}
 PSD=\frac{Q_l-Q_s}{Q_l}
\end{equation}
The $Q_l$ is the energy integral of a whole pulse. $Q_s$ is the energy integral of the front edge of the pulse. Different fall time of neutron signals and $\gamma$-ray signals in the EJ-301 detector make this method feasible.
Neutron signals have longer attenuation time than $\gamma$-ray signals in the neutron detector. The peak in Fig.\ref{neutron detector} with higher psd are the neutron signals.

\begin{figure}
  \centering
  \includegraphics[scale=0.35]{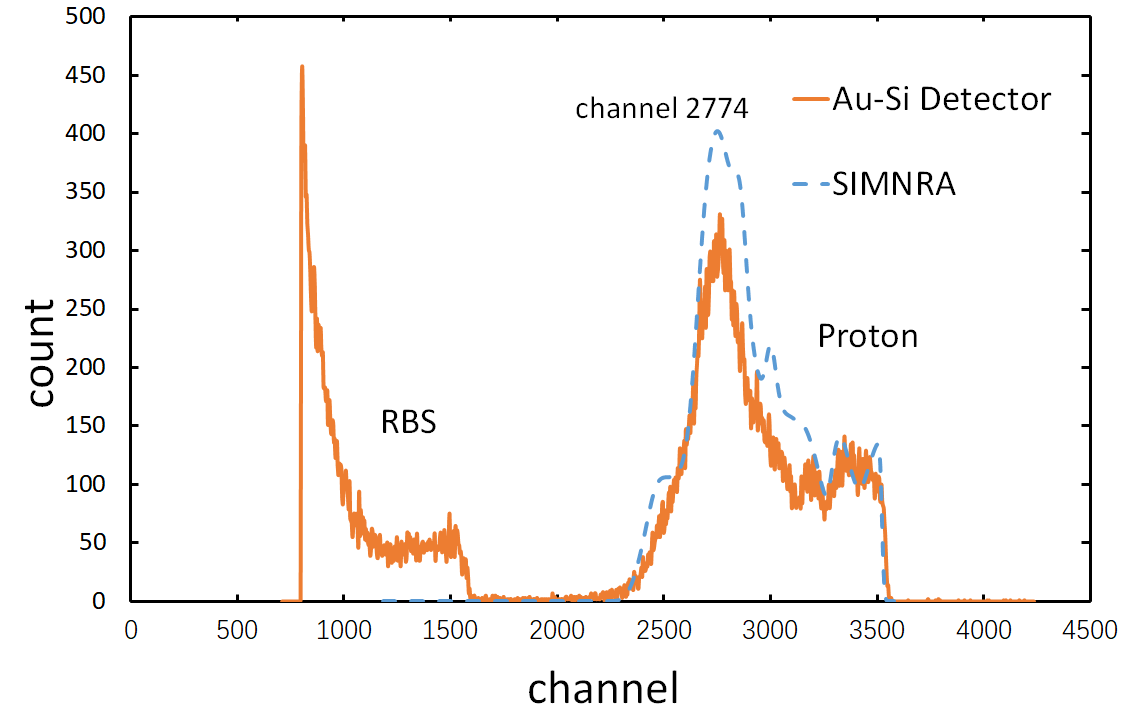}\\
  \caption{The signals of Au-Si detector at $135\degree$, incident energy 2 MeV and the signals simulated by SIMNRA \cite{mayer1997simnra}.}
  \label{proton detector}
\end{figure}

Fig.\ref{proton detector} shows an example of the proton signals and the results simulated by SIMNRA at $135\degree$ with the incident energy 2 MeV. The Q value of this reaction is about 2.7 MeV which helps to identify and calibrate the proton from the detector. The signals of which channel below 1600 are the Rutherford backscattered(RBS) deuterons recoiled from the target. The signals of which channel above 2000 are the protons. The proton's energy distribution is caused by the energy change of deuteron beam in the thick target. The peak at channel 2774 are the signals of protons when the incident deuteron energy is 2 MeV. The signals between channel 2774 to channel 3555 belong to the protons produced with the incident energy below 2 MeV due to the ionization in the thick target. The energy shift is caused by the center of mass motion. The Au-Si detector is put at $135\degree$ in the opposite direction to beam in the center of mass frame. With this condition, the higher the incident energy of the deuteron, the faster the center of mass moves which leads to a lower speed of outgoing proton in the backward direction.

\begin{figure}

  \includegraphics[scale=0.6]{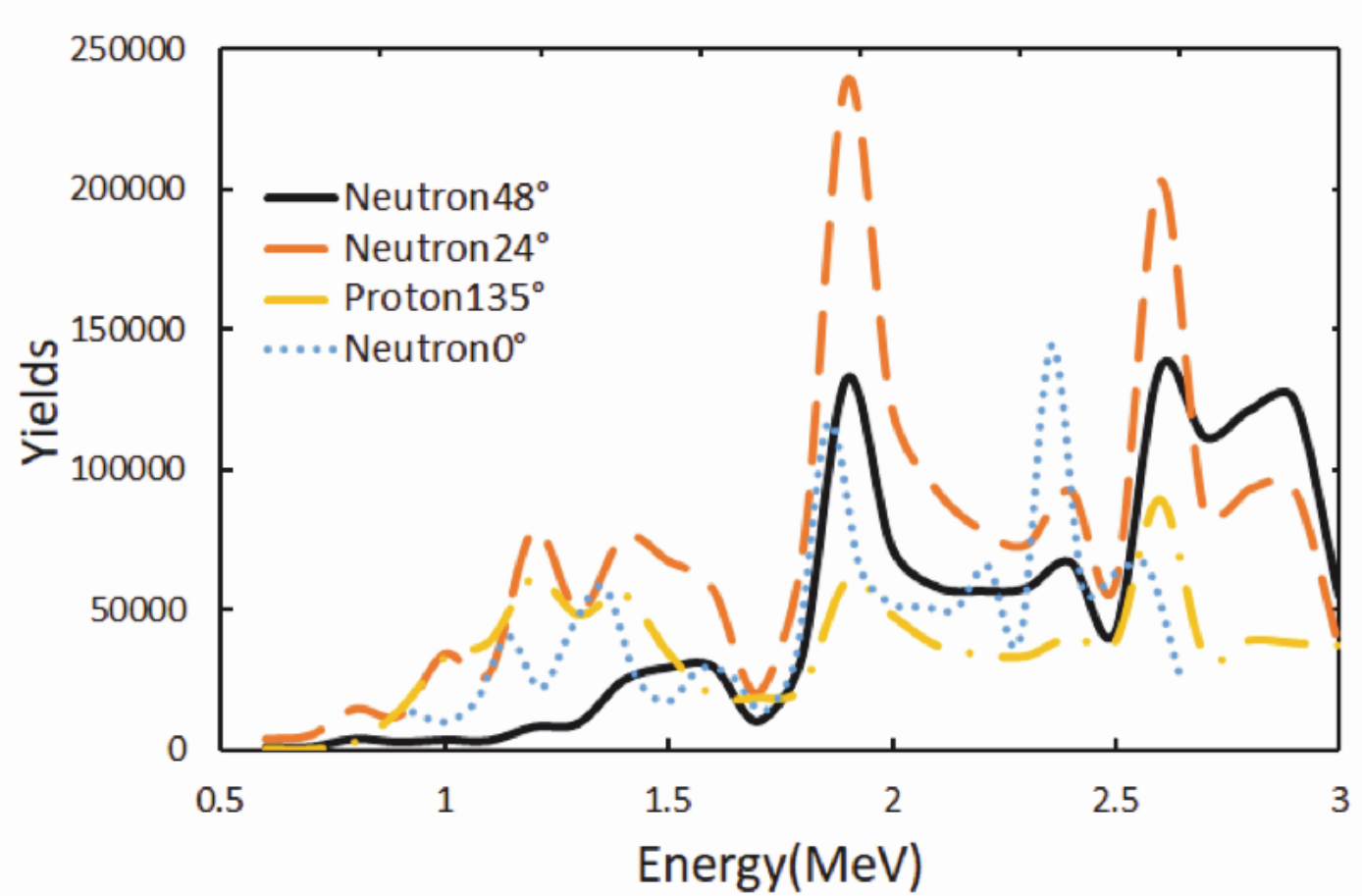}\\
  \caption{The yield of neutron at $0\degree$, $24\degree$, $48\degree$ and the yield of proton at $135\degree$ from 0.6 MeV to 3 MeV.}
  \label{yield}
\end{figure}

Fig.\ref{yield} shows the yield of the neutrons and protons when the incident energies vary from 0.6 MeV to 3 MeV. The yield is corrected by excluding the background and convert into 300 seconds. The errors of the yields depend on the statistical magnitude. Since every counts in each measurement is above 560, the errors are below 4.2\%. Due to the sparking of the Faraday-cup and the imprecise incident beam intensity, the extrem vaule in the plot can't give any convincing conclusion about cross section of the reactions. Our method enable us to exclude the error in the incident beam and find the physical information in this two reactions.
\begin{figure}

  \includegraphics[scale=0.55]{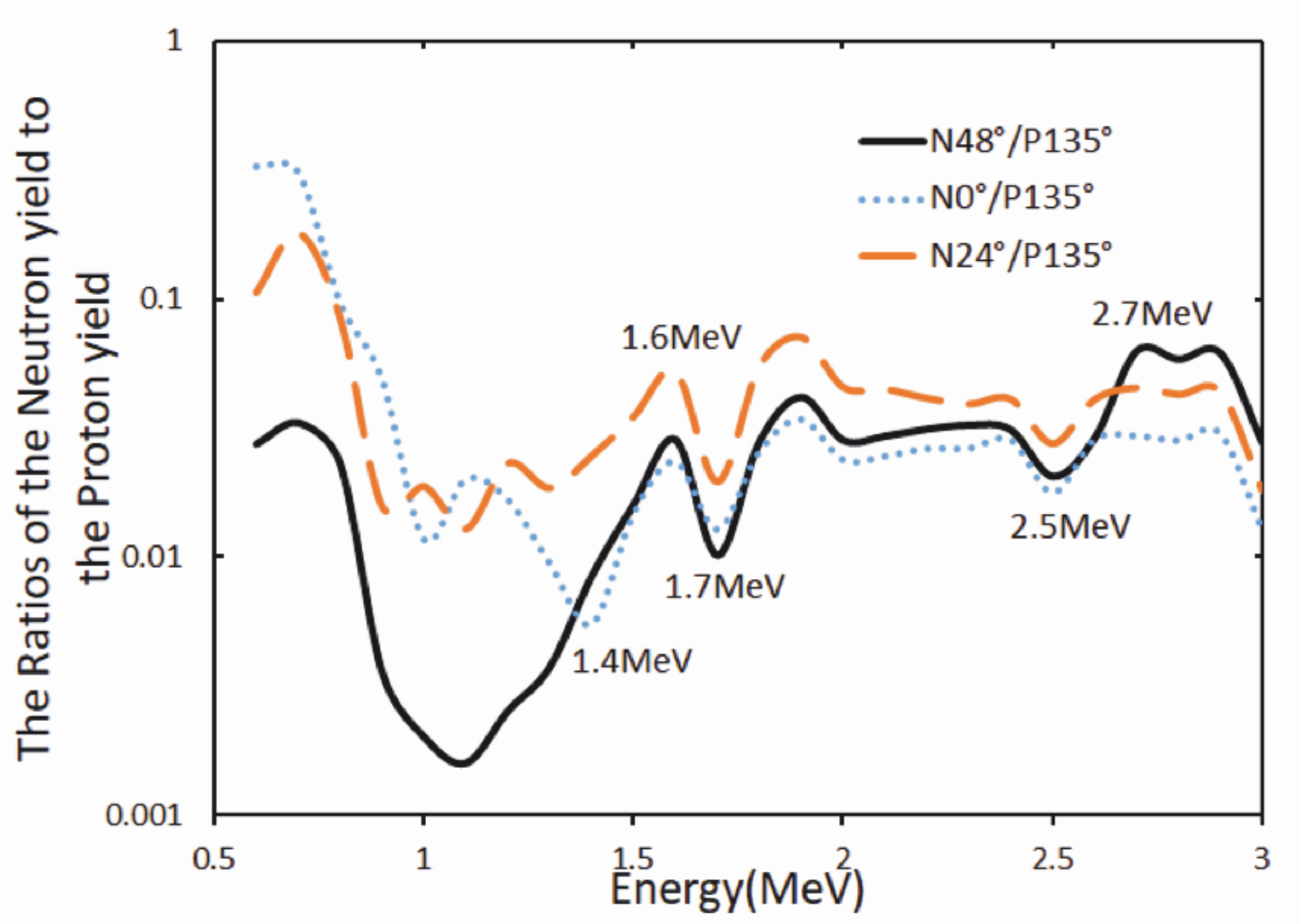}\\
  \caption{The ratio of neutron yield to the proton yield in our experiment from 0.6 MeV to 3 MeV. }
  \label{ratio}
\end{figure}

\section{discussion}
\label{sec:discussion}
The target used in this work is a nature graphite which contains $^{12}C$ and $^{13}C$. The proton and neutron channels from the reaction between $^{13}C$ and deuteron must be considered. 
The cross section of the $^{13}C(d,n)^{14}$N reaction is about two to three of the cross section of the $^{12}C(d,n)^{13}N$ reaction from the ref. \cite{michelnn_excitation_nodate,brune_total_1992}. The abundance of $^{12}C$ is 98.89\% while the abundance of $^{13}C$ is 1.109\%. Although the $^{13}C$ has bigger cross section in this experiment than that of the $^{12}C$, the great difference in the abundance make the neutron yield in the $^{13}C$ target, within 3.3\% of the total neutron yield,  insignificant.

Other possible reaction channels in the experiment such as $^{12}C(d,n+p)^{12}C$, $^{12}C(d,n+d)^{11}C$ and $^{12}C(d,n+\alpha)^{9}B$ should also be considered. The data from the TENDL-2017 Nuclear data library \cite{tendl2017} shows that the cross-section of these reaction from 0.6 to 3 MeV are too small to be taken into account. The dominant reactions in this experiment are  $^{12}C(d,n)^{13}N$ and $^{12}C(d,p)^{13}$C.

In the Fig.\ref{ratio}, the ratio is different in three neutron detection directions due to the angular distribution, but it seems not a problem in our analysis. When the incident energy gives rise to a resonance in the reaction, the yield of the reaction will increase equally in all directions. The ratio of yield is still recognizable.

The ratio between neutron and proton yields presents some hints of the resonance in $^{12}C(d,n)^{13}$N and $^{12}C(d,p)^{13}$C. The maximum points and the minimum points in the curves in Fig.\ref{ratio} may show some features which present evidence for the complexity of these nuclear reactions.

\begin{figure}
  \includegraphics[scale=0.55]{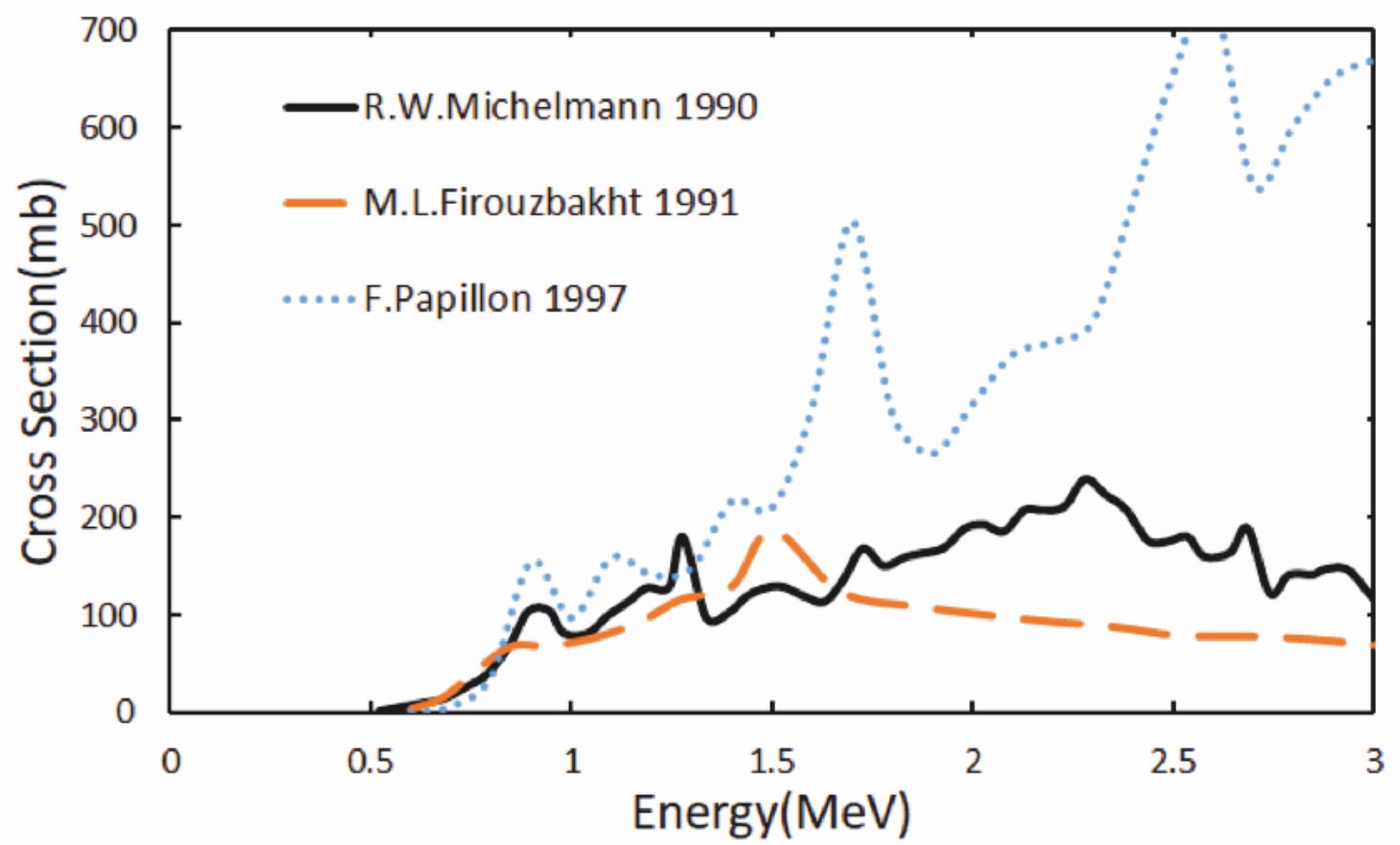}\\
  \caption{The cross section of neutron from the ref.\cite{firouzbakht_cross-section_1991,michelnn_excitation_nodate}. The cross section of proton at 170$\degree$ from the ref.\cite{papillon_analytical_1997}.}
  \label{reference cross section}
\end{figure}

The minimum points in our curves may be either the effect of a sudden jump in proton yield or a sudden drop in neutron yield as we have discussed in Methods part. Thus, if the cross section of neutron doesn't show decrease at the resonance points, the cross section of proton must reach to a maximum value, which provide a hint for the resonance state for the proton channel.
The fig.\ref{reference cross section} shows the cross section of neutron and cross section of proton from ref.\cite{firouzbakht_cross-section_1991,michelnn_excitation_nodate,papillon_analytical_1997}. The plot shows whether the cross of section neutron decreases at the possible resonance energies or not.  With these data, we can check the possible resoances in the Fig.\ref{ratio}. The proton detection in the ref.\cite{papillon_analytical_1997} is measured at $170\degree$ which is different from our detection but since the angular distribution shows no effect on the resonance recognition, ref.\cite{papillon_analytical_1997} still works in following analysis.

For $^{12}C(d,p)^{13}$C reaction, two resonances are found at 1.7 MeV and 2.5 MeV in all directions and one resonance is found at 1.4 MeV in the 0$\degree$ direction. In the Fig.\ref{reference cross section}, the cross section of neutron doesn't show decrease at 1.4 MeV, 1.7 MeV or 2.5 MeV which garantees the resonance points for the proton channel. The one at 1.7 MeV comes from the shell in the $^{14}$N with energy 11.76 MeV \cite{csedreki_measurements_2014,papillon_analytical_1997}. The one at 2.5 MeV is from the shell in the $^{14}$N with energy 12.4 MeV \cite{papillon_analytical_1997}. The one at 1.4 MeV is also consists with the ref.\cite{papillon_analytical_1997} coming from the shell in the $^{14}$N whose energy 11.51 MeV, spin and parity is 3$^{+}$. 

In the same way, for $^{12}C(d,n)^{13}$N reaction, resonances are found at 2.7 MeV in the 0$\degree$ and 24$\degree$. Two other resonance are found at 1.6 MeV and 1.9 MeV in all directions. In the Fig.\ref{reference cross section}, the cross section of proton doesn't show increase at 1.6 MeV and 2.7 MeV, which confirms the 1.6 MeV resonance point in $^{12}C(d,n)^{13}$N while the cross section at 1.9 MeV show big decrease meaning this minimum vaule may not be a reasonance. The resonance at 2.7 MeV
 consist with the ref.\cite{jaszczak_c_1969,michelnn_excitation_nodate} coming from the excitation state of the $^{14}$N with energy 12.58 MeV. The one at 1.6 MeV is consistent with the result in the ref.\cite{bonner_cross_1956}. The resonance at 1.6 MeV in the $^{12}C(d,n)^{13}$N reaction comes from the shell in the $^{14}$N of which excitation energy, spin and parity are 11.66 MeV, 2$^{+}$ or 3$^{+}$. 

Table 2 shows the summary of the resonances in the $^{12}C(d,n)^{13}$N and in the $^{12}C(d,p)^{13}$C from the ref.\cite{michelnn_excitation_nodate,jaszczak_c_1969,bonner_cross_1956,firouzbakht_cross-section_1991,csedreki_measurements_2014,papillon_analytical_1997} and ours. 

\begin{figure*}
 \includegraphics[scale=0.6]{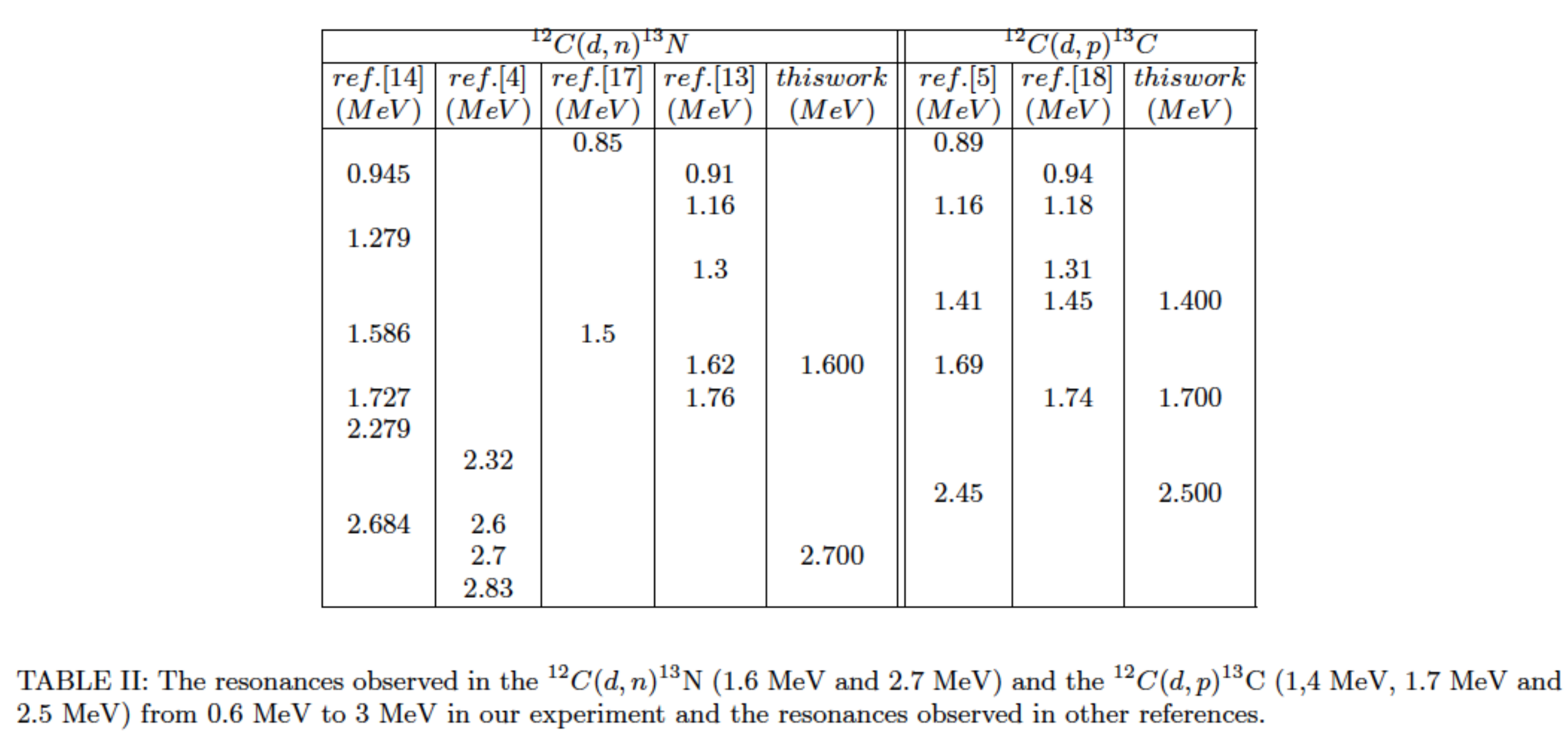}\\
  \label{reference cross section}
\end{figure*}

In the Fig.\ref{ratio}, it is found the ratio changes rapidly in the lower energy part, but with the energy increasing, it relatively becomes more stable and harder for us to recognize the resonance points. Though the yield change rate reflects the cross section, the higher the energy of incident particle in the thick target, the more the energy change the deuterons go through in the whole bombarding process. That leads to a more significant difference between cross section and ratio of yield. In this circumstance, only those great resonance points can be found in our curves. This is also the reason that the resonances we found are less than that in other authors' works.

\section{Summary}
\label{sec:conc}
By performing and studying of the bombardment of deuteron on thick carbon target, the resonance of the two reaction in the bombardment, $^{12}C(d,n)^{13}$N and  $^{12}C(d,p)^{13}$C, is observed. The resonances when the incident deuteron energies are 1.4 MeV, 1.7 MeV and 2.5 MeV in the $^{12}C(d,p)^{13}$C reaction are affirmed. The resonances  when the incident deuteron energies are 1.6 MeV and 2.7 MeV in the $^{12}C(d,n)^{13}$N reaction are affirmed. 
We suggest the ratio of neutron yield to proton one as a new way to study the resonances in the $^{12}C(d,n)^{13}$N reaction and $^{12}C(d,p)^{13}$C reaction.
This method eliminates the error not only from the incident beam intensity, but also from the subtle change in the target during the bombard by making use of the two reactions. 

There is still some space for the improvement in the near future. On one hand, though both neutron resonance and proton resonance points are found, the number of each are relatively small. On the other hand, this method's feasibility strongly depends on the incident energy. The resonance may be hard to recognize when the incident energy is too high or the incident energy range is too wide.
If necessary, the experiment can be improved by the way as followed. Firstly, a thin target thick about 1 $\mu m$ is recommended to take the place of the thick target. The error from the change of deuteron energy in the thick target can be eliminated then. Secondly, the event by event method can be applied to get rid of the background signals. This method can be realized by putting a detector to detect the incident beam and make a coincidence measurement among the incident particle and the outgoing particles. However, as a first step, our method guarantees a low system error and gives a hint for more precise cross section measurements.

\medskip

 This work was partially supported by the Strategic Priority Re-search Program of the Chinese Academy of Sciences (Grant No. XDB16 and XDPB09), the National Natural Science Foundation of China under Contract Nos. 11890714 and 11421505,  the Key Research Program of Frontier Sciences of the CAS under Grant No. QYZDJ-SSW-SLH002.  We thank  Z. T. He from Shanghai Institute of Applied Physics, Chinese Academy of Sciences, for the target preparation in this work.
 

\end{document}